# A spatiotemporal analysis of participatory sensing data "tweets" and extreme climate events toward real-time urban risk management

Yoshiki Yamagata, Daisuke Murakami, Gareth W. Peters and Tomoko Matsui

## Abstract

Real-time urban climate monitoring provides useful information that can be utilized to help monitor and adapt to extreme events, including urban heatwaves. Typical approaches to the monitoring of climate data include the acquisition of weather station monitoring and also remote sensing via satellite sensors. However, climate monitoring stations are very often distributed spatially in a sparse manner, and consequently, this has a significant impact on the ability to reveal exposure risks due to extreme climates at an intra-urban scale (e.g., street level). Additionally, such traditional remote sensing data sources are typically not received and analyzed in real-time which is often required for adaptive urban management of climate extremes, such as sudden heatwaves. Fortunately, recent social media, such as Twitter, furnishes real-time and high-resolution spatial information that might be useful for climate condition estimation.

The objective of this study is utilizing geo-tagged tweets (participatory sensing data) for urban temperature analysis. We first detect tweets relating hotness (hot-tweets). Then, we study relationships between monitored temperatures and hot-tweets via a statistical model framework based on copula modelling methods. We demonstrate that there are strong relationships between "hot-tweets" and temperatures recorded at an intra-urban scale, that we reveal in our analysis of Tokyo city and its suburbs. Subsequently, we then investigate the application of "hot-tweets" informing spatio-temporal Gaussian process interpolation of temperatures as an application example of "hot-tweets". We utilize a combination of spatially sparse weather monitoring sensor data, infrequently available MODIS remote sensing data and spatially and temporally dense lower quality geo-tagged twitter data. Here, a spatial best linear unbiased estimation (S-BLUE) technique is applied. The result suggests that tweets provide some useful auxiliary information for urban climate assessment. Lastly, effectiveness of tweets toward a real-time urban risk management is discussed based on the analysis of the results.

---

Y. Yamagata • D. Murakami
Center for Global Environmental Research, National Institute for Environmental Studies, Tsukuba, Japan
Email: yamagata@nies.go.jp

D. Murakami (Corresponding author)
Email: murakami.daisuke@nies.go.jp

G. W. Peters
Department of Statistical Science, University College London, London, U.K.
Email: gareth.peters@ucl.ac.uk

T. Matsui U.K.
Department of Statistical Modeling, Institute of Statistical Mathematics, Tachikawa, Japan
Email: tmatsui@ism.go.jp

# 1. Introduction

The ability to develop management responses and adaptive strategies to manage extreme weather related events, such as urban heatwaves, is a major concern that is actively studied in the urban literature (e.g., Nakamichi et al., 2013; Yamagata and Seya, 2013). A key challenge faced in such adaptive management approaches is the ability to accurately monitor in real time the intra-urban scale climate conditions (e.g., district level). However, monitoring stations that assess the climate conditions are very often allocated sparsely in space, and it is usually difficult or impossible to analyze district level temperatures and humidity using the resulting data obtained from such sensor networks due to the sparsity of locations at an intra-urban scale. This is a known problem discussed in papers such as Zhang (2010) and Nevat et al. (2013). To illustrate this point for the urban environment in Tokyo, Figure 1 displays the spatial distribution of temperature monitoring stations of Japan Metrological Agency in Tokyo prefecture. The number of the monitoring stations is only 8, and it would be impossible to analyze district level temperatures using data at the monitoring stations only.

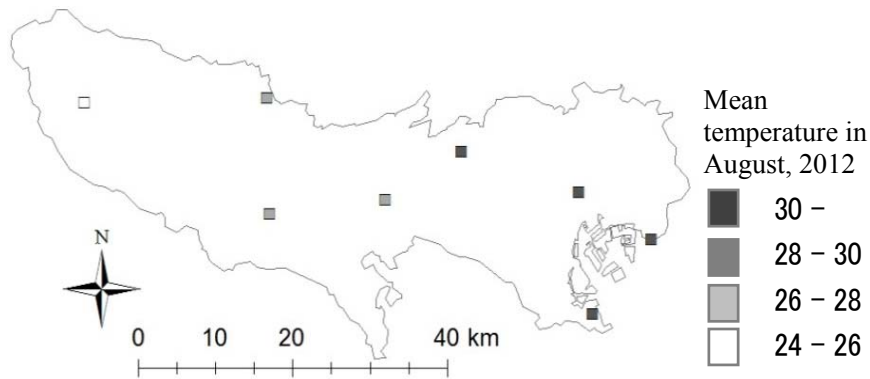

**Fig. 1.** Spatial distribution of climate monitoring stations in Tokyo.

In addition to data from weather monitoring stations, often remote sensing satellite data provides another popular approach for climate condition estimation. For example, satellite imageries from MODIS (Moderate resolution imaging spectroradiometer) reveal ground surface temperatures with spatial resolution of 1km. However, we cannot acquire real-time climate information from such satellite images 24 hours a day; in the case of Ja-pan, MODIS imagery is available only at four time points, which are around AM11:30, PM1:30, PM8:30, and PM10:30 each day.

Given the challenges faced by sparsity (in space) of ground based weather and climate monitoring stations and the sparsity (in time) of remote sensing data to monitor in real time intra-urban climates, we seek a novel alternative approach that can supplement these accurate data sources with a less accurate but prevalent in time and space lower quality data source. Hence, to provide real-time district-level climate information for adaptive public urban management of extreme climate events, we need to explore a novel solution to supplement the missing data in regions far from monitoring sensor sites (see Figure 1). We will demonstrate in this paper how to begin to utilize Twitter (http://twitter.com) as an auxiliary information source to help to explain and extrapolate in an informed manner the local urban climate, especially, in this study the local intra-urban temperatures of Tokyo city. Twitter is a social media that allows uses to post short messages (up to 140 characters in Japanese) called "tweets." Since a large amount of tweets are posted at arbitrary times and locations, we investigate whether they may provide real-time and high spatial resolution information relating to temperature and in particular extreme temperature events such as heatwaves. It is unknown as to what extent such data would be useful in helping to resolve this issue of estimation in the presence of scarce data from real climate monitoring sensor networks. Therefore, we aim to study the utility of this additional Twitter data in forming estimation of temperature in local urban environments to supplement the results obtained from sensor monitoring station recordings.

Application of tweets to such estimation problems is a recently active topic, for instance Thelwall et al. (2011) show that major events significantly change sentiment of tweets. In Bollen et al. (2011) they utilized twitter data to estimate public mood, and then applied the results to financial market analysis. Of most relevance to the study we undertake, it was demonstrated recently in An et al. (2014) that tweets change depending on climate events, i.e. the information content in tweets and the topic of tweets is causally affected by climate conditions of the tweeter. They also show that tweets **concerning climate change are increasing**. However, most twitter studies including these studies just mentioned ignore the spatial dimension and structure of the data, this is probably due to the difficulty of obtaining geotagged data on such tweet data records.

It is starting to be demonstrated that the spatial data associated with tweets may also be significant and informative when utilizing twitter data. For instance, some studies suggest significant influence of location on tweets, see for example, Hahmann et al. (2014) who show that intensity of tweets about facilities (Airport, Bakery, Cinema,...) change depending on the distance to the facility, while Li et al. (2012) demonstrate that tweets effectively reveal important information relating to local events such as car accidents and earthquakes. Tweets might also effectively describe local urban climate, it is this premise that we aim to investigate in this study.

As a first step to utilizing tweets in urban climate monitoring, we first analyze statistical relationships between tweets and temperatures of their corresponding locations at intra-urban resolutions. In particular we focus on unpleasant hotness and heatwaves as the target of our extreme climate study. The subsequent sections of the manuscript are organized as follows. Section 2 discusses how to exploit information pertaining to hotness from the twitter data. Section 3 analyzes relationship between tweets and temperatures. Section 4 demonstrates an application example of tweets for local urban temperature estimation. Finally, section 5 concludes our discussion.

## 2. Temperature-related data

### 2.1 Description of Twitter data

We acquired data for tweets which included the following important attributes: the text message; a user id code; the spatial coordinates; and the time of tweeting. Such Twitter databases are known as geo-tagged tweets, the particular data set we utilized is commercially available and was collected by Nightley Inc. and purchased for this research project. The dataset consists of tweets posted in Tokyo between August 1 to August 31 in 2012. As a privacy policy, this data does not include all of the recorded tweets posted but instead only accounts for approximately 1% of a randomly sampled set of the geo-tagged tweets. Following, An et al. (2014), we exclude "re-tweets," which involves data that comes as a consequence of a re-posting of someone else's tweet. The reason for this choice is that it is difficult to detect sentiment from a re-tweet. The resulting sample size we utilized involves 130,332 geo-tagged tweets. In Figure 2, we provide a spatial scatter plot of their locations throughout the period under study, we see that the tweets tend to be distributed densely across Tokyo which provides a good resolution of information in the intra-urban Tokyo districts. The dense coverage is a marked contrast with the sparse distribution of temperature monitoring stations, which is shown in Figure 1. The dense tweeter information can now be assessed to see if it would be valuable to complement temperature information at unmonitored sites.

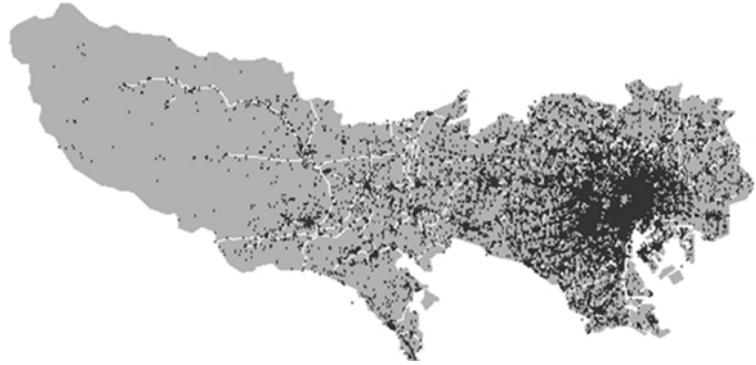

**Fig. 2.** Spatial distribution of the tweeter data. White lines represent railways

**Table 1.** Extracted synonyms of "Hot" and/or "Hot + Uncomfortable"

| Japanese | English | Japanese | English |
|---|---|---|---|
| あつい | Hot | 蒸し | Muggy |
| 熱い | Hot | 水分補給 | Rehydration |
| 暑 | Heat | 体調管理 | Health management |
| 猛暑 | Heat wave | 猛烈に | Furious |
| 炎天下 | Blazing sun | だるい | Dull |
| 真夏日 | Hot day | 死ぬ | Dying |
| 残暑 | Lingering summer heat | 異常 | Abnormal |
| 熱中症 | Heat illness | 不快感 | Discomfort |
| バテ | Summer heat | 不快 | Discomfort |
| 寝苦しい | Cannot sleep well | イヤ | Unpleasant |
| 夏本番 | Midsummer | 嫌 | Unpleasant |
| 日差し | Sunlight | クソ | Shit |
| 照り | Reflected heat | Orz | Discouragement |
| 湿度 | Humid | きつい | Hard |
| 湿気 | Moisture | 辛い | Hard |
| 汗 | Sweat | 大変 | Hard |
| ジメジメ | Damp | しんどい | Tired |
| ムシムシ | Humid | 厳しい | Severe |
| ベタベタ | Sticky | 苦手 | Weak |

## 2.2 Extraction of temperature related hotness tweets: "hot-tweets"

In this study we extracted only tweets that specifically related to unpleasant hotness using an approach known as "keygram," (see http://kizasi.jp/labo/keygram/keygram.py/) which utilizes a Japanese synonyms dictionary to associated words (characters) to words of hotness (heat related). Table 1 summarizes synonyms of "hot" or "hot + un comfortable", which were extracted through the use of keygram. Indeed, they are likely to explain unpleasant hotness. We use a dummy variable indicating 1 if the tweets included any of the hotness-related keywords, and 0 otherwise, which we then refer to as "hot-tweets".

Spatial distribution of hot-tweets are distributed in Figure 3. This figure suggests that hot-tweets increase in nearby station areas and that they increase in the 2nd and 3rd week, which are the middle of the summer in Japan, and decrease in the 4th week, which is the end of the summer.

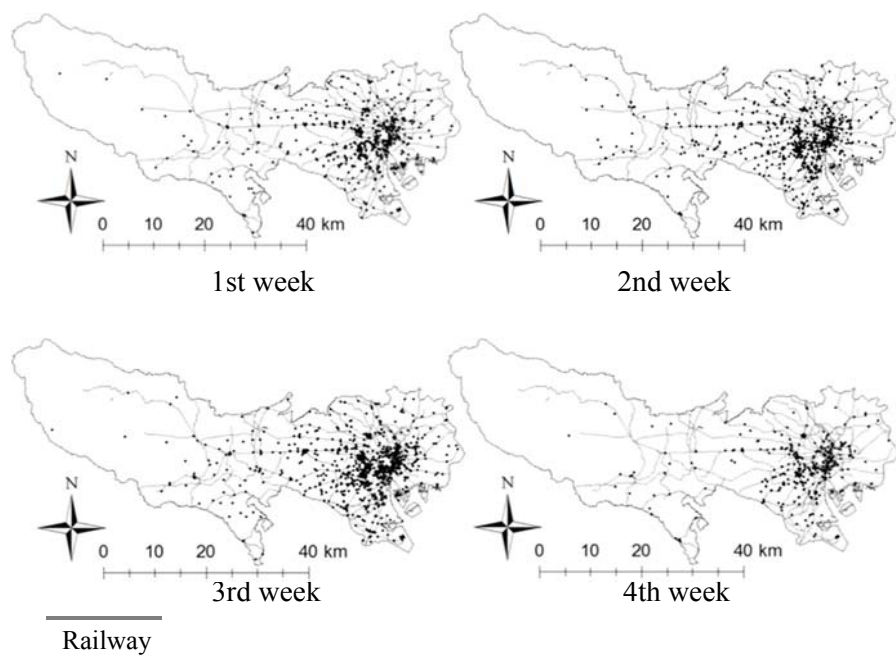

**Fig. 3.** Spatial distribution of hot-tweets for each week in August, 2012

## 3. Analysis of tweets and temperature over time

### 3.1 A regression analysis based on generalized additive logistic model

To analyze factors that explain "hot-tweets", we apply a logistic additive regression model (e.g., Wood, 2006), which describes binary variables by a weighted sum of smoothing (nonparametric) terms as well as parametric terms. Suppose that $p_i = P(hot\text{-}tweet_i = 1)$, where $i \in \{1,...N\}$ is an index of tweet, then, our applied logistic additive model describes $p_i$ according to the following model structure:

$$\mathrm{logit}(p_i) = \log\left(\frac{p_i}{1-p_i}\right) = \alpha + \sum_p x_{i,p}\beta_p + s1(day_i) + s1(time_i) + s2(lon_i, lat_i) \quad (1)$$

where $\alpha$ is a constant, $x_{i,p}$ is an explanatory variable, and $\beta_p$ is the coefficient, $day_i$ and $hour_i$ are the day and the time (in minutes) that $i$-th tweet is posted, respectively, and $lon_i$ and $lat_i$ are longitude and latitude of the tweeted location.

**Table 2.** Explanatory variables

| Variables | Description |
|---|---|
| Const. | Constant (intercept); |
| Tokyo dist. | Log. of the minimum railway distance between the nearest station and either of the principal main rail stations in Tokyo (Tokyo, Shinjuku, Ikebukuro, Shibuya, and Shinagawa stations); |
| Station dist. | Log. of the distance to the nearest railway station; |
| Day pop. | Daytime population density; |
| Night pop. | Nighttime population density; |
| Park dist. | Log. of the distance to the nearest urban park; |
| River dist. | Log. of the distance to the nearest river; |
| Commerce | Dummy of commercial district versus non-commercial (residential or other); |
| Industry | Dummy of industrial district |

Table 2 summarizes explanatory variables. $s_1(\cdot)$ is the smooth spline function that models non-linear impacts of *day$_i$* and *time$_i$*. For the smoothing function, we used the conventional thin plate spline (Wood, 2003; 2006). $s_2(\cdot)$ is the bivariate spatial smoothing spline function. Here, we use the Tensor product smoothing operator for $s_2(\cdot)$ (Wood et al., 2013). Eq.(1) outperforms some other density estimation methods (e.g., the *K*-function-based method) in that multi-level (i.e., daily and hourly) temporal patterns that can be considered. The parameter estimation can be performed using function for general additive models such as gamm in the mgcv package in R, which we used in this analysis.

Figure 4 shows the estimated non-linear impacts. $s_1(day_i)$ shows a time trend that illustrates that "hot-tweets" have a tendency to decrease around August 10 of 2012 and then to increase after this date. To clarify the reason for this behavior, we considered the daily mean temperatures which are plotted in the left side of Figure 5. In agreement, with these findings from the hot-tweets, we also see that the mean temperatures decrease around August 10, 2012. Thus, we postulate that perhaps the intensity of hot-tweets may reflect basic trends in daily temperatures. Estimated $s_1(time_i)$ shows a tendency for hot-tweets to increase in the morning, and decrease at night, such diurnal patterns are expected in local intra-urban climates. The intensity of hot-tweets is low at night and the same as the observed temperatures (see the middle of Figure 5). However, the increase of hot-tweets in the morning is somewhat inconsistent with the trend of temperatures, whose high peak is typically in the afternoon. On the other hand, change of hourly temperatures, which is plotted in the right subplot of Figure 5, has the high peak in the morning and the low peak at night, which is in agreement with the intensity of hot-tweets. Thus, we postulate that perhaps "hot-tweets" may also reflect changes of temperature rather than absolute temperatures themselves. Estimated $s_2(lon_i, lat_j)$ shows a spatial pattern that hot-tweets increase in areas around train stations in the suburbs of Tokyo, which is the middle of this figure, and decrease in the western non-urban area and the eastern central Tokyo area. Monitored temperatures (see, Figure 1) do not display such tendency; hot-tweets might explain spatial pattern of unmonitored temperatures.

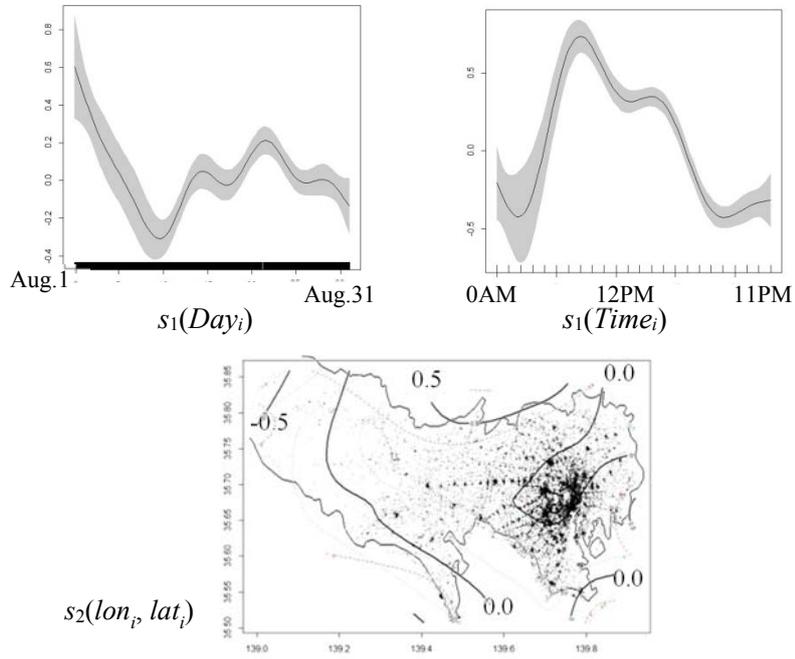

**Fig. 4.** Estimated non-linear impacts. Black lines show the estimated impacts, and gray area indicate their 95% confidential intervals

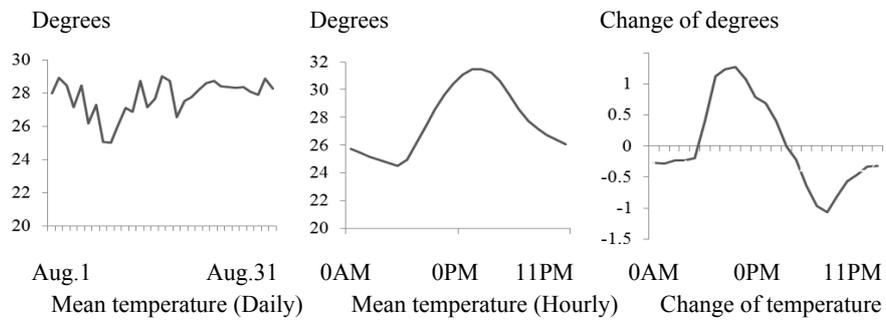

**Fig. 5.** Mean temperatures observed at monitoring stations (left: daily means; middle: hourly means; right: hourly mean changes of temperatures).

**Table 3.** Parameter estimates

| Variables | Estimates | z-value | Significance |
|---|---|---|---|
| Const. | -6.17 | -14.4 | *** |
| Tokyo dist. | $2.44 \times 10^{-3}$ | 0.17 | |
| Station dist. | $-7.81 \times 10^{-3}$ | -0.43 | |
| Day pop. | $3.69 \times 10^{-7}$ | 0.16 | |
| Night pop. | $-1.99 \times 10^{-7}$ | -0.50 | |
| Park dist. | $-4.11 \times 10^{-1}$ | -4.79 | *** |
| River dist. | $-1.89 \times 10^{-1}$ | -4.15 | *** |
| Commerce | $-1.28 \times 10^{-1}$ | -2.67 | ** |
| Industry | $-1.09 \times 10^{-1}$ | -1.56 | |

. ** and * represents significant levels of 5% and 1% respectively

Table 3 summarizes the estimates of $β_p$ which demonstrate through regression analysis based assessment in what types of urban area people will tend to feel unpleasant hotness and be driven to comment about such local climate conditions. The results from this analysis show that, while indicators of urbanization, including Tokyo dist., Station dist., Day pop., and Night pop., are statistically significant, indicators of natural environment, including Park dist. and River dist. are negatively significant at the 1% level. Therefore, these results suggest that "hot-tweets" tend to decrease in areas collocated around new parks and/or rivers, this may be due to the following reasons: the cooling effect of green and water areas which are known to act as heat-sinks in local urban climates; as well as the amenities available in many of these areas, which is also likely to reduce the chance of feeling unpleasant hotness and therefore tweeting about this unpleasant micro climate. Commerce districts are also found to be negatively significant at the 5% level, which means that people in commercial areas tend to experience less unpleasantness due to heat in such intra-urban environments. Again, speculatively this may be due to amenities and air-conditioned office and leisure spaces in such locations, also resulting in a reduction the effect of unpleasant heat in such areas.

### 3.2 A correlation and extremal dependence analysis

As Hahmann et al. (2014) mentioned the "*application of tweet-analyses for high resolution applications should be approached with care as the corre-*

*lation between the contents and the locations of tweets that is required for these applications is probably often too low.*" Hence, a study of the correlation between tweets and temperature of their corresponding locations would be an important starting point to utilize tweets for high resolution urban climate analysis. We also argue that extremal measures of dependence and concordance are also required to be considered in this spatial-temporal setting, in this regard we also include an analysis of spatial extreme dependence relationships captured through the notion of extremal tail dependence.

To analyze the correlation, the twitter data must be associated with the temperature data, whose monitoring locations (see, Figure 1) are incompatible with tweeted locations (see, Figure 2). Besides, since values of hot-tweets (1 or 0) can be very noisy, it would be preferable to utilize the underlying process of generating hot-tweets, which is more stable, rather than hot-tweet values. Thus, we utilize the probability of hot-tweets = 1, i.e. *p*(hot-tweet=1) for the correlation analysis. This is because *p*(hot-tweet=1) can be considered as hot-tweets after a noise removal.

The correlation between *p*(hot-tweet=1) and temperatures is 0.30, while the correlation between *p*(hot-tweet) and change of temperatures is 0.46. Thus, the change in temperature has a greater linear association with the probability of "hot-tweets". We plot *p*(hot-tweet=1) against temperatures (left side of Figure 6) and change of temperatures (right side of Figure 6). Apparently, lower temperatures seem to have stronger positive correlation with *p*(hot-tweet=1) than higher temperatures. Also, greater positive change of temperature seems to have stronger correlation with *p*(hot-tweet=1) than the other positive or negative change of temperature. In other words, the correlation between *p*(hot-tweet=1)s and temperature/change of temperature is asymmetric. Unfortunately, such an asymmetric correlation structure cannot be captured by the standard correlation coefficient.

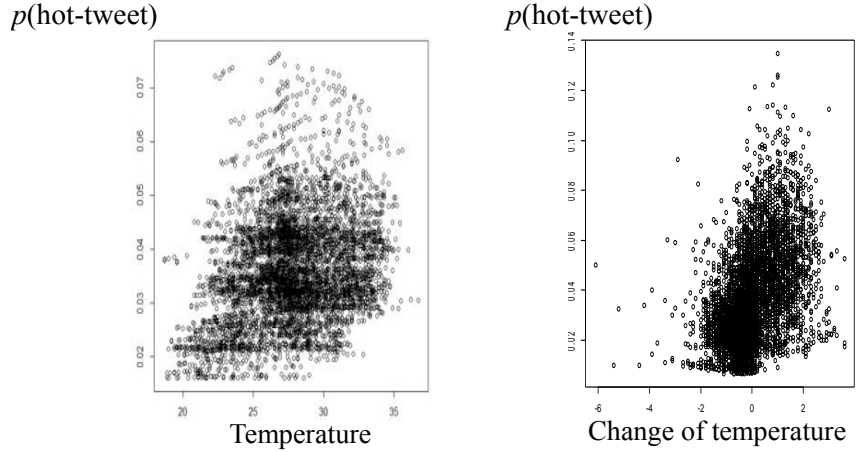

**Fig. 6.** Plots of dHtweets against temperatures and change of temperatures

To accommodate the potential for other measures of concordance as well as asymmetric correlation features such as those just analyzed we advocate a statistical modelling framework known as copula modelling. Copulas are a distributional parametric model based framework that provides functional model specifications that can accommodate a variety of different notions of concordance and dependence such as asymmetric correlation structures observed in this data analysis, see a detailed review of copula modelling in Cruz et al. 2014 (chapters 7-10). The widespread use of copula models is based on the results obtained from Sklar's theorem (Sklar, 1959), which assures the existence of a unique probability distribution function called a copula, $C(\cdot)$, which satisfies the following relationship:

$$F(x_1, \cdots x_p) = C(F_1(x_1), \cdots F(x_p)), \qquad (2)$$

where $F(x_1, \ldots x_p)$ is a $p$-dimensional cumulative distribution function (CDF) of random variables, $x_1, \ldots x_p$, and $F_1(x_1), \ldots F_p(x_p)$ are their continuous marginal distribution functions. As such we see that according to Sklar's theorem the copula is a CDF that joints these marginal distributions together and therefore directly captures the dependence and associations between each random variable, independent of the marginal scaling behavior.

After differentiation one obtains the general expression for the joint probability density function (PDF) of random vector $x_1, \ldots x_p$, denoted by $f(x_1, \ldots x_p)$, according to the following decomposition based on Eq.(2) as follows:

$$f(x_1, \cdots x_p) = c(F_1(x_1), \cdots F(x_p)) \prod_{j=1}^{p} f_j(x_j), \qquad (3)$$

where $f_j(x_j)$ is the marginal PDF of $x_j$. Eq.(2) implies that $f(x_1,... x_p)$ can be decomposed into mutually independent elements, $f_j(x_j)$, and the element that joints them, the copula density function $c(F(x_1),... F(x_p))$. Our interest is in the copula density $c(F(x_1),... F(x_p))$, which can be selected to describe the correlation between temperatures or change of temperatures and $p$(hot-tweet). Many different copula models are available to be considered, see discussions in Cruz et al. 2014 (chapters 7-10) and the references therein.

While $c(F(x_1),... F(x_p))$ can be modeled by both parametric or non-parametric approaches, we apply the latter, which is more flexible. We use a penalized hierarchical B-spline-based copula estimation approach of Schell and Ruppert (2013). This approach models copula using Eq.(4):

$$c(u_1, \cdots u_p) \approx \sum_{k=1}^{K} b_k \phi_k(u_1, \cdots u_p), \qquad (4)$$

where $u_1,... F(x_1),... u_p = F(x_p)$, and $\phi_k(u_1, \cdots u_p)$ is $k$-th B-spline basis function, and $b_k$ is the corresponding coefficient. See, Schell and Ruppert (2013), for more details on properties of such a semi-parametric copula model.

In Figure 7 we plot the estimated copula model in Eq.(4) for the relationship between temperature and p(hot-tweet). The result suggests the significant correlation between lower temperatures and $p$(hot-tweet) and in addition perhaps even a lower tail dependence feature. In other words, variations of low temperatures, e.g. at night are explained well by $p$(hot-tweet). Strong correlation is also found between greater positive change of temperature and $p$(hot-tweet). That is, rapid temperature increase, e.g. in morning are explained well by $p$(hot-tweet).

As an additional analysis of such relationships we also consider alternative concordance measures such as extremal tail dependence. Upper and lower tail dependences are quantified non-parametrically, as detailed in Ames et al. (2014), via the following concordance measure for upper tail dependence:

$$\lambda_u = 2 - \min\left[2, \frac{\log \hat{C}\left(\frac{n-k}{n}, \frac{n-k}{n}\right)}{\log\left(\frac{n-k}{n}\right)}\right], \qquad (5)$$

where the empirical copula is given by
$\hat{C}\left(\frac{n-k}{n}, \frac{n-k}{n}\right) = \frac{1}{n}\sum_{i=1}^{n} 1\left(\frac{R_{1i}}{n} \leq u_1, \frac{R_{2i}}{n} \leq u_2\right)$, and $R_{ji}$ is the rank of the variable in its marginal dimension that makes up the pseudo data (data transformed via the marginal distribution functions to the unit hyper-cube). $n$ is the sample size. $k$ is set as the 1st, 2nd,... 20th percentiles to estimate the lower tail dependences, and as 80st, 81nd,... 99th percentiles to estimate the upper tail dependences.

Upper tail dependences estimated in each percentile, $\lambda_u$, are averaged and summarized in Table 4. Lower tail dependences are summarized in the same table as well. This table shows the marked lower tail dependence between temperature and $p$(hot-tweets=1). In contrast, any extremal upper tail dependence was not found between temperature and $p$(hot-tweets=1). This table also shows moderate upper and lower tail dependences between change of temperatures and $p$(hot-tweets=1).

In summary, while $p$(hot-tweets=1) explain both temperature and change of temperature, $p$(hot-tweets) have stronger relationship with the latter. Besides, $p$(hot-tweets=1) have moderate upper and lower tail dependences with change of temperatures as well. From this dependence analysis we conclude that there may be some utility in utilizing geo-tagged twitter data to help estimate and forecast temperature change in a high-resolution manner, in order to supplement data from sparsely located monitoring stations and infrequent remote sensing data sets. Furthermore, the correlation between $p$(hot-tweet=1) and temperatures suggests that perhaps tweets may carry some information that could be useful for estimation and forecasts of the magnitude of temperatures in intra-urban environments.

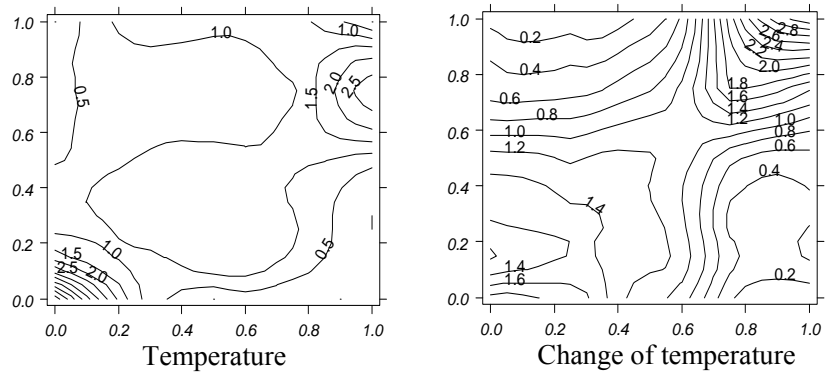

**Fig. 7.** Non-parametric copula estimates

**Table 4.** Lower and upper tail dependence estimates

|  | Lower tail | Upper tail |
|---|---|---|
| Temperature | 0.40 | 0.00 |
| Change of temperature | 0.20 | 0.23 |

## 4. Estimation of intra-urban spatial resolution temperature profiles using "hot-tweet" geo-tagged twitter data

Based on the features observed in the geo-tagged "hot-tweet" data we now undertake an application that utilizes this data to perform intra-urban temperature profile analysis. Hence, in this part of the analysis we utilize "hot-tweet" data as an auxiliary information to capture local temperature interpolation and combine it with more accurate sparse weather/climate monitoring data as well as remote sensing MODIS data for Tokyo prefecture. To achieve this we must first develop a spatial-temporal temperature interpolation model considering both monitored temperatures and "hot-tweet" data. Since "hot-tweet" data occurs at any time and at any place, the idea is that they should provide greater spatial coverage for spatial temperature profile reconstruction in areas without any local sensor monitoring stations. Therefore, the aim of this section is to assess the utility of such geo-tagged "hot-tweet" data in informing such spatial temperature map reconstruction for intra-urban scale analysis.

### 4.1 Spatial temperature field reconstruction at intra-urban resolution

*4.1.1 Model for monitored temperatures*

The model we develop is based on a spatial-temporal Gaussian process in which we express the de-trended temperature at the *i*-th monitoring station, whose spatiotemporal coordinates are denoted by $\mathbf{x}_i$, $y(\mathbf{x}_i)$, as follows

$$y(\mathbf{x}_i) = f(\mathbf{x}_i) + v(\mathbf{x}_i), \quad v(\mathbf{x}_i) \sim N(0, \sigma^2), \tag{6}$$

where $\sigma^2$ is a variance parameter. $f(\mathbf{x}_i)$ is a Gaussian process, which we express as follows:

$$f(\mathbf{x}_i) \sim N(0, c(\mathbf{x}_i, \mathbf{x}_j)), \tag{7}$$

where, and $c(\mathbf{x}_i, \mathbf{x}_j) = E[f(\mathbf{x}_i)f(\mathbf{x}_j)]$ is a covariance function.

If $c(\mathbf{x}_i, \mathbf{x}_j)$ is given by a kernel function of the distance $d(\mathbf{x}_i, \mathbf{x}_j)$ and the time-lag $t(\mathbf{x}_i, \mathbf{x}_j)$ between $\mathbf{x}_i$ and $\mathbf{x}_j$, the resulting Gaussian process describes both spatial and temporal dependencies. Here, one has to pay particular attention to the form of the kernel in terms of how space and time features may interact in the temperature profile reconstruction. In this analysis we adopt a product-sum model proposed in (De Iaco et al., 2001) which provides a popular spatiotemporal modelling specification which is formulated as follows according to the spatiotemporal covariance kernel functions:

$$c(\mathbf{x}_i, \mathbf{x}_j) = \sigma_s^2 k[d(\mathbf{x}_i, \mathbf{x}_j)] + \sigma_t^2 k[t(\mathbf{x}_i, \mathbf{x}_j)] + \sigma_{st}^2 k[d(\mathbf{x}_i, \mathbf{x}_j)] k[t(\mathbf{x}_i, \mathbf{x}_j)], \quad (8)$$

where $\sigma_s^2$, $\sigma_t^2$, $\sigma_{st}^2$ are unknown variance parameters. The spatial and temporal kernel functions $k[d(\mathbf{x}_i, \mathbf{x}_j)]$ and
$k[t(\mathbf{x}_i, \mathbf{x}_j)]$ in Eq.(8) can be defined using exponential function as follows:

$$k[d(\mathbf{x}_i, \mathbf{x}_j)] = \exp\left[-\frac{d(\mathbf{x}_i, \mathbf{x}_j)}{r_s}\right], \quad (9)$$

$$k[t(\mathbf{x}_i, \mathbf{x}_j)] = \exp\left[-\frac{t(\mathbf{x}_i, \mathbf{x}_j)}{r_t}\right], \quad (10)$$

where $r_s$ and $r_t$ are unknown range parameters of spatial and temporal dependencies, respectively.

Use of a basis function representation is an alternative approach to construct the spatiotemporal covariance function $c(\mathbf{x}_i, \mathbf{x}_j)$. In such a setting one would suppose that $\mathbf{w}_s(\mathbf{x}_i)$, $\mathbf{w}_t(\mathbf{x}_i)$, and $\mathbf{w}_{st}(\mathbf{x}_i)$ are vectors of spatial and temporal basis function values at location $\mathbf{x}_i$, then, $c(\mathbf{x}_i, \mathbf{x}_j)$ can be modeled as follows:

$$c(\mathbf{x}_i, \mathbf{x}_j) = \mathbf{b}'_s \mathbf{w}_s(\mathbf{x}_i)\mathbf{w}_s(\mathbf{x}_j)' \mathbf{b}_s + \mathbf{b}'_t \mathbf{w}_t(\mathbf{x}_i)\mathbf{w}_t(\mathbf{x}_j)' \mathbf{b}_t + \mathbf{b}'_{st} \mathbf{w}_{st}(\mathbf{x}_i)\mathbf{w}_{st}(\mathbf{x}_j)' \mathbf{b}_{st}, \quad (11)$$

where $\mathbf{b}_s$, $\mathbf{b}_t$, and $\mathbf{b}_{st}$ are coefficient vectors. Although Eq.(11) is a rank 1 covariance model, by assuming $\sigma^2 > 0$, the resulting covariance matrix of $y(\mathbf{x}_i)$ is always positive definite (see Eq.6). Following Eqs.(9) and (10), we defined the basis functions by the following exponential functions:

$$w_s(\mathbf{x}_i) = \exp\left[-\frac{d(\mathbf{x}_i, \widetilde{\mathbf{x}}_a)}{\widetilde{r}_s}\right], \quad (12)$$

$$w_t(\mathbf{x}_i) = \exp\left[-\frac{t(\mathbf{x}_i, \widetilde{\mathbf{x}}_a)}{\widetilde{r}_t}\right], \quad (13)$$

$$w_{st}(\mathbf{x}_i) = \exp\left[-\frac{d(\mathbf{x}_i, \widetilde{\mathbf{x}}_a)}{\widetilde{r}_s}\right] \exp\left[-\frac{t(\mathbf{x}_i, \widetilde{\mathbf{x}}_a)}{\widetilde{r}_t}\right], \tag{14}$$

where $w_s(\mathbf{x}_i)$, $w_t(\mathbf{x}_i)$, and $w_{st}(\mathbf{x}_i)$ are $i$-th element of $\mathbf{w}_s(\mathbf{x}_i)$, $\mathbf{w}_t(\mathbf{x}_i)$, and $\mathbf{w}_{st}(\mathbf{x}_i)$, respectively. Here, we denote by $\widetilde{\mathbf{x}}_a$, the spatiotemporal coordinates of the $a$-th anchor point, which is placed anywhere in the target period in the target area, and $\widetilde{r}_s$ and $\widetilde{r}_t$ are known range parameters. Eqs.(12), (13), and (14) generate the $a$-th radial basis function based on the spatial and/or the temporal distance separating $\mathbf{x}_i$ and $\widetilde{\mathbf{x}}_a$. When anchor points are densely separated in space, many basis functions are generated (the number of basis functions equals to the number of anchor points), and the resulting process model is usually more accurate than the model with fewer basis functions. In contrast, models with fewer basis functions are usually computationally more efficient. Hence, the number of anchor points must be selected judiciously. Following Cressie and Johannesson (2008), who discuss a basis function-based spatial interpolation, the known range parameters $\widetilde{r}_s$ and $\widetilde{r}_t$ are given by 1.5 times of the shortest distance and time-lag between anchor points, respectively.

While the former kernel-based approach has extensively been discussed (e.g., Cressie, 1993; Cressie and Wikle, 2013; Nevat et al. 2015), discussions of the basis function approach is still limited. However, the basis function approach is the computational more efficient: it is in nature a dimension reduction approach, and the computational complexity can be $O(K^3)$, where $K$ is the number of basis functions (see, e.g., Cressie and Johannesson, 2008) whereas the computational complexity of the kernel approach is $O(N^3)$, where $N$ is the sample size. Section 4.2 compares effectiveness of the two approaches by a Monte Carlo simulation.

### 4.1.2 Model for "hot-tweet" geo-tagged twitter data

Suppose that $y_{ht}(\mathbf{x}_I)$ is a dummy variable of indicating 1 if the tweet at location $\mathbf{x}_I$ is a hot-tweet and 0 otherwise. We describe the probability of $y_{ht}(\mathbf{x}_I) = 1$ using Eqs.(15) and (16), which are defined as

$$P(y_{ht}(\mathbf{x}_I) = 1 | f(\mathbf{x}_I)) = \sum_{p=0}^{1} P(y_{ht}(\mathbf{x}_I) = 1 | y_{ht\_0}(\mathbf{x}_I) = p) P(y_{ht\_0}(\mathbf{x}_I) = p | f(\mathbf{x}_I)), \tag{15}$$

$$y_{ht\_0}(\mathbf{x}_I) = \begin{cases} 1 & if \quad f(\mathbf{x}_I) \geq T \\ 0 & otherwise \end{cases}, \tag{16}$$

where $T$ is a pre-determined threshold. $y_{ht\_0}(\mathbf{x}_I)$ is a dummy variable indicating 1 if $f(\mathbf{x}_I)$ exceeds the threshold temperature $T$, and 0 otherwise. Eq.(15) assumes that one tends to comment about hotness when $f(\mathbf{x}_I)$ exceeds $T$ (i.e., $y_{ht}(\mathbf{x}_I)$ tends to be 1 if $y_{ht\_0}(\mathbf{x}_I) = 1$).

### *4.1.3 Efficient spatial-temporal temperature field reconstructions for intra-urban resolutions*

In this section we develop a spatial linear estimator for temperature data based on the S-BLUE framework proposed in (Nevat et al, 2015) and (Peters et al, 2015). The aim of this analysis is to interpolate the temperature at a new spatial locations $\mathbf{x}_*$, at which no measurement data is available, by estimating the underlying spatiotemporal process of temperatures by minimizing the mean squared error (MSE), which is formulated as follows:

$$E[(f(\mathbf{x}_*) - \hat{f}(\mathbf{x}_*))^2], \qquad (17)$$

where $f(\mathbf{x}_*)$ is the true unknown process and $\hat{f}(\mathbf{x}_*)$ is the estimate. For simplicity, we denote $f(\mathbf{x}_*)$ as $f_*$ hereafter. We select a class of spatial temperature field estimators, $\hat{f}_*$, that satisfy the following linear functional form given as

$$\hat{f}_* = \hat{\alpha} + \hat{\mathbf{B}}\mathbf{Y} = \arg\min_{\alpha,\mathbf{B}} E\left[(f_* - (\alpha + \mathbf{B}\mathbf{Y}))^2\right], \qquad (18)$$

where $\hat{\alpha} \in R$, $\hat{\mathbf{B}} \in R^{1 \times N}$, and $\mathbf{Y}$ is a vector observations that stacks vectors of $y(\mathbf{x}_i)$ and $y_{ht}(\mathbf{x}_I)$, which we denote $\mathbf{y}$ and $\mathbf{y}_{ht}$, respectively (i.e., $\mathbf{Y} = [\mathbf{y}', \mathbf{y}'_{ht}]'$). Nevat et al. (2015) show that the minimization of the MSE under Eq.(18) yields the following estimator of $f_*$, which we call the spatial best linear unbiased estimator (S-BLUE):

$$\hat{f}_* = E[f_*\mathbf{Y}']E^{-1}[\mathbf{Y}\mathbf{Y}']\mathbf{Y}. \qquad (19)$$

The elements in $E[f_*\mathbf{Y}']$ and $E[\mathbf{Y}\mathbf{Y}']$ are given based on Eqs.(6) and (7), as follows:

$$E[f_*\mathbf{Y}'] = E[f_*\mathbf{y}', f_*\mathbf{y}'_{ht}] = [\mathbf{c}'_* \quad E[f_*\mathbf{y}'_{ht}]], \qquad (20)$$

$$E[\mathbf{Y}\mathbf{Y}'] = E\begin{bmatrix} \mathbf{y}\mathbf{y}' & \mathbf{y}\mathbf{y}'_{ht} \\ \mathbf{y}_{ht}\mathbf{y}' & \mathbf{y}_{ht}\mathbf{y}'_{ht} \end{bmatrix} = \begin{bmatrix} \sigma^2\mathbf{I} + \mathbf{C} & E[\mathbf{y}\mathbf{y}'_{ht}] \\ E[\mathbf{y}_{ht}\mathbf{y}'] & E[\mathbf{y}_{ht}\mathbf{y}'_{ht}] \end{bmatrix}, \qquad (21)$$

where $\mathbf{I}$ is an identity matrix, $\mathbf{c}_*$ is a column vector whose $i$-th element is $c(\mathbf{x}_i, \mathbf{x}_*)$, and $\mathbf{C}$ is a matrix whose $(i, j)$-th element is $c(\mathbf{x}_i, \mathbf{x}_j)$. In this study,

$c(\mathbf{x}_i, \mathbf{x}_*)$ and $c(\mathbf{x}_i, \mathbf{x}_j)$ are given by either of the kernel function-based Eq.(8) or the basis function-based Eq.(11). In addition to the point estimator for the S-BLUE in Eq. (19) one may also define the accuracy of this estimator, as shown in Eq. (22). The associated MSE of the S-BLUE is given by

$$\sigma_*^2 = c(\mathbf{x}_*, \mathbf{x}_*) - E[f_*\mathbf{Y}']E^{-1}[\mathbf{YY}']\mathbf{Y}. \qquad (22)$$

A summary of the algorithmic steps required to calculate this S-BLUE estimator is provided in Algorithm 1.

---

**Algorithm 1**: S-BLUE field reconstruction

Input : $\mathbf{Y}, \mathbf{x}_i, \mathbf{x}_*, \sigma^2, \mathbf{C}, \mathbf{c}$

Output : $\hat{f}_*$

1: Calculate the cross-correlation vector $E[f_*\mathbf{Y}']$

2: Calculate the covariance matrix $E[\mathbf{YY}']$

3: Calculate the S-BLUE of the intensity of the spatial field at a location $\mathbf{x}_*$, as follows:

$$\hat{f}_* = E[f_*\mathbf{Y}']E^{-1}[\mathbf{YY}']\mathbf{Y}$$

---

In summary, we developed models for monitoring temperatures and hot-tweets in section 4.1.1 and section 4.1.2, respectively, and developed the S-BLUE, Eq.(19), which interpolates temperatures at $\mathbf{x}_*$ considering both monitored temperatures and hot-tweets. The remaining components to discuss for this spatial estimator that incorporates remote sensing data, ground based monitoring data and geo-tagged twitter data involves specification of how best to evaluate or estimate the spatial conditional moments given by $E[f_*\mathbf{y}'_{ht}]$, $E[\mathbf{yy}'_{ht}]$, and $E[\mathbf{y}_{ht}\mathbf{y}'_{ht}]$ in Eqs.(20), (21), which we will discuss in section 4.1.3, and how to estimate parameters in $c(\mathbf{x}_i, \mathbf{x}_*)$ and $c(\mathbf{x}_i, \mathbf{x}_j)$, which we will discuss in section 4.1.4.

### 4.1.3 Approximation of the cross-product moment terms in the S-BLUE estimator

The detailed derivations for the estimators of these spatial cross correlations is provided in Peters and Matsui (2015), here we present a summary of these results as they pertain to the application of these techniques in this paper. We consider the $I$-th element of $E[f_*\mathbf{y}'_{ht}]$, the $(i, J)$-th element of $E[\mathbf{yy}'_{ht}]$, and the $(I, J)$-th element of $E[\mathbf{y}_{ht}\mathbf{y}'_{ht}]$, which are in Eqs.(20), (21), are expressed as follows in Lemma 1 and Lemma 2 (see detailed derivations in Peters and Matsui, 2015).

*Lemma 1 (Cross-Correlation between Spatial Process and Observations)*

The *I*-th element $E[f_*y_I]$ of $E[f_*\mathbf{y}'_{ht}]$, which describes a cross-correlation between spatial processes at *I*-th tweeted site and *-th prediction site, is given by

$$E[f_*y_I] = -\frac{c(\mathbf{x}_*,\mathbf{x}_I)}{c(\mathbf{x}_I,\mathbf{x}_I)}\sum_{k=0}^{1}\Pr(y_I=1|y_{I\_0}=k)c(\mathbf{x}_I,\mathbf{x}_I)$$
$$\{\phi(T_{k+1};0,\sigma^2+c(\mathbf{x}_I,\mathbf{x}_I))-\phi(T_k;0,\sigma^2+c(\mathbf{x}_I,\mathbf{x}_I))\}, \quad (23)$$

where $T_0 = -\infty$, $T_1 = T$, and $T_2 = \infty$ (see, Eq.16), and $\phi(T_k;0,\sigma^2+c(\mathbf{x}_I,\mathbf{x}_I))$ is the value of the probability density function of the normal variable with mean 0 and variance $\sigma^2 + c(\mathbf{x}_I, \mathbf{x}_I)$ at $T_k$.

*Lemma 2 (Cross-Correlation between Observations)*

The (*i*, *I*)-th element of $E[\mathbf{yy}'_{ht}]$, $E[y_iy_I]$, which describes a cross-correlation between spatial processes at *i*-th monitoring station and at *I*-th tweeted site, is given by

$$E[y_iy_J] = \frac{c(\mathbf{x}_i,\mathbf{x}_J)}{c(\mathbf{x}_J,\mathbf{x}_J)}$$
$$\sum_{k=0}^{1}\Pr(y_J=1|y_{J\_0}=k)E[f_J\{\Phi(T_{k+1};f_J,\sigma^2)-\Phi(T_k;f_J,\sigma^2)\}], \quad (24)$$

The (*I*, *J*)-th element of $E[\mathbf{y}_{ht}\mathbf{y}'_{ht}]$, $E[y_Iy_J]$, which describes a cross-correlation between spatial process values at *I*-th and *J*-th tweeted sites, is given by

$$E[y_Iy_J] = \sum_{k=0}^{1}\sum_{l=0}^{1}\Pr(y_I=1|y_{I\_0}=k)\Pr(y_J=1|y_{J\_0}=k)$$
$$E[\{\Phi(T_{k+1};f_I,\sigma^2)-\Phi(T_k;f_I,\sigma^2)\}\{\Phi(T_{k+1};f_J,\sigma^2)-\Phi(T_k;f_J,\sigma^2)\}], \quad (25)$$

where $\Phi(\lambda_k;f_i,\sigma^2) = \int_{-\infty}^{T_k}\phi(T;f_i,\sigma^2)dT$ is the value of the cumulative distribution function of $\phi(T_k;f_i,\sigma^2)$ at $T_k$.

Eqs.(23), (24), and (25) are derived by assuming $\mu(\cdot)$, which appears in Peters and Matsui (2015), equals 0. This assumption is imposed because we assume temperatures after detrending (e.g., using a linear regression model: see section 4.1.1). Also, the number categories in the binary variables, *L*, which is in Peters and Matsui (2015), is given by 2 following our assumption of using hot-tweets, which take 0 or 1.

Fortunately, Lemma 1 provides a closed form solution of $E[f_*y_I]$. Accordingly, $E[f_*\mathbf{y}'_{ht}]$, which is a part in the S-BLUE equation is readily calculated once covariance functions, $c(\mathbf{x}_i,\mathbf{x}_j)$, are estimated. In contrast, solutions of $E[y_iy_I]$ and $E[y_Iy_J]$ include expectation terms. To calculates the expectations, we need to evaluate $\int_{-\infty}^{T_k}\phi(T;f_i,\sigma^2)dT$ and $f_I$. If $f_i$ and $\sigma^2$ are known, they have closed form solutions. The end of subsection 4.1.4 discusses how to estimate them in a computationally efficient manner, which is important to implement S-BLUE considering large data such as twitter data.

### 4.1.4 Estimation

This section discusses how to estimate parameters in $c(\mathbf{x}_i, \mathbf{x}_j)$ (and $c(\mathbf{x}_i, \mathbf{x}_*)$), which are contained in the expressions for $E[f_*\mathbf{y}'_{ht}]$, $E[\mathbf{yy}'_{ht}]$, and $E[\mathbf{y}_{ht}\mathbf{y}'_{ht}]$, and S-BLUE equation (19) as well. Note that we use monitored temperatures only in the parameter estimation step.

The kernel function (Eq.8)-based $c(\mathbf{x}_i, \mathbf{x}_j)$ can be estimated using the weighted least squares (WLS)-based method (e.g., Cressie, 1993), whose spatiotemporal version is recently installed in gstat, which is a standard geostatistical package in R (http://www.r-project.org/). We apply the WLS-based method to estimate the product-sum kernel function, Eq.(8).

On the other hand, we apply an expectation-maximization (EM) algorithm of Peters and Matsui (2015) which extended the formulation of Hoff and Niu (2012) to the mixed sensor fusion spatial covariance estimation context. We utilize this approach of Peters and Matsui (2015) in order to estimate the basis function-based $c(\mathbf{x}_i, \mathbf{x}_j)$, Eq.(11). We briefly overview the key components of this approach and explain the algorithm. We first substitute the basis function-based $c(\mathbf{x}_i, \mathbf{x}_j)$, Eq.(8), in to the Gaussian process model, Eq.(6). The resulting model is expressed by a matrix notation as follows:

$$\mathbf{y}=\mathbf{f}+\mathbf{v}, \quad \mathbf{f}\sim N(\mathbf{0},\mathbf{Bww'B'}), \quad \mathbf{v}\sim N(\mathbf{0},\sigma^2\mathbf{I}), \quad (25)$$

where $\mathbf{f}$ describes the Gaussian process and $\mathbf{v}$ the white noise. Eq.(25) is identical to the following equation:

$$\mathbf{y}=\mathbf{Bw\Gamma}+\mathbf{v}, \quad \mathbf{\Gamma}\sim N(\mathbf{0},\mathbf{I}), \quad \mathbf{v}\sim N(\mathbf{0},\sigma^2\mathbf{I}), \quad (26)$$

where $\mathbf{B}$ is a matrix of coefficients. The log-likelihood of this model yields

$$l(\sigma^2, \mathbf{B}, \mathbf{E}, \mathbf{W}) = c - \frac{1}{2}\sum_{i=1}^{n}\log\left|\sigma^2\mathbf{I} + \mathbf{B}\mathbf{w}_i\mathbf{w}_i'\mathbf{B}'\right| -$$

$$\frac{1}{2}\sum_{i=1}^{n}tr\left[(\sigma^2\mathbf{I} + \mathbf{B}\mathbf{w}_i\mathbf{w}_i'\mathbf{B}')^{-1}y(\mathbf{x}_i)^2\right], \tag{27}$$

where $\mathbf{w}_i = [\mathbf{w}_s(\mathbf{x}_i)', \mathbf{w}_t(\mathbf{x}_i)', \mathbf{w}_{st}(\mathbf{x}_i)']'$.

The EM-algorithm identifies the maximum likelihood (ML) estimates of $\sigma^2$ and $\mathbf{B}$. To establish the algorithm, we first consider the conditional distribution of the random effects nuisance parameter, $\Gamma_i$, which is the $i$-th element of $\Gamma$, which can be shown to obtain the conditional distribution as follows

$$\Gamma_i \mid \hat{\sigma}^2, \hat{\mathbf{B}} \sim N(m_i, v_i), \tag{28}$$

$$v_i = \left[1 + \frac{1}{\sigma^2}\mathbf{w}_i'\mathbf{B}'\mathbf{B}\mathbf{w}_i\right]^{-1} \quad m_i = \frac{v_i}{\sigma^2}y(\mathbf{x}_i)\mathbf{B}\mathbf{w}_i$$

Using Eq.(28), the conditional ML estimators of $\mathbf{B}$ and $\sigma^2$ are derived by differentiating the log-likelihood to obtain the expressions given as follows

$$\hat{\mathbf{B}} = \tilde{\mathbf{E}}'\tilde{\mathbf{W}}(\tilde{\mathbf{W}}'\tilde{\mathbf{W}})^{-1}, \tag{29}$$

$$\hat{\sigma}^2 = \frac{1}{n}(\tilde{\mathbf{E}} - \tilde{\mathbf{W}}\hat{\mathbf{B}})'(\tilde{\mathbf{E}} - \tilde{\mathbf{W}}\hat{\mathbf{B}}), \tag{30}$$

where $\tilde{\mathbf{W}}$ is a $2n \times q$ matrix with $i$-th row given by $m_i\mathbf{w}_i$ and $(n+i)$-th row by $s_i\mathbf{w}_i$, and $\tilde{\mathbf{E}}$ is a $2n \times p$ matrix of the residuals given by $[\mathbf{E}', \mathbf{0}]'$, where $\mathbf{0}$ is a matrix of zeros with its dimension equals to $\mathbf{E}'$, a detailed discussion of the derivation steps and re-parameterization of this model to obtain Eq. (29) and (30) are provided in Peters and Matsui (2015).

Now, based on Eqs.(28)-(30), we can develop an EM-algorithm-based ML estimation whose procedure is summarized as follows:

1- Initialize $\sigma^2$ and $\mathbf{B}$
2- Estimate $m_i$ and $v_i$ using Eq.(28)
3- Estimate $\sigma^2$ and $\mathbf{B}$ using Eq.(29), (30)
4- Iterate the steps (ii) and (iii) until the estimates of $\sigma^2$ and $\mathbf{B}$ converge

Fortunately, Eqs.(29) and (30), which calculate least squared estimates, and Eq.(28) are computationally efficient. Hence, this estimation approach is likely to be a computationally efficient alternative to the standard kernel-based estimation, which can be highly computationally inefficient when spatiotemporal data are modeled.

After estimating **B**, the value of the Gaussian process at $x_i$, $f_i$, can be estimated by $Bw_i$ [1](see Eq.25). Likewise, the value of the Gaussian process at a tweeted site $x_I$ can be estimated by $Bw_I$ where $w_I = [w_s(x_I)', w_t(x_I)', w_{st}(x_I)']'$. This study applies $f_i$ and $f_I$, which are estimated like that, to estimate $\int_{-\infty}^{T_k} \phi(T; f_i, \sigma^2) dT$ and $f_I$, which are in $E[y_i y_I]$ and $E[y_I y_J]$ (see the previous subsection). In other words, after the EM-algorithm-based parameter estimation, we obtain closed form solutions of $E[y_i y_I]$ and $E[y_I y_J]$, and $E[f_* y_I]$, and the S-BLUE Eq.(19) as well. This approach is computationally efficient, and, accordingly, would be useful for S-BLUE considering a large dataset such as twitter data.

### 4.2. Monte Carlo simulation comparing performance of kernel based model with basis function based model.

#### *4.2.1 Assumptions on the accuracy and efficiency study*

Before going to the S-BLUE-based temperature interpolation, this section performs a Monte Carlo simulation that compares effectiveness of the kernel function-based approach, which uses Eq.(8), and the basis function-based approach, which uses Eq.(11), in terms of model accuracy and computational time.

Sample sites are distributed in a two-dimensional space. Their *X* and *Y* coordinates are determined by scaling random samples from *N*(0,1) to make the minimum and the maximum coordinates 0.0 and 1.0, respectively (see Figure 8). Synthetic data are generated for the *N* sample sites and 30 by 30 regular points, whose maximum and minimum coordinates are 0.1 and 0.9, respectively. The synthetic data is generated using the Gaussian process whose expectation equals 0 and covariance function is exp(-$d(x_i, x_j)$/0.4). Figure 8 illustrates sample sites, the regular points, and a resulting simulated synthetic Gaussian process that is used to perform the Monte Carlo comparative study.

In this simulation study, data generated in the *N* sample sites are used for model estimation, the data generated in the 30 by 30 points are used to evaluate predictive accuracy of models. We demonstrate two types of sim-ulations. The first simulation examines the influence of sample size on predictive accuracy. The simulation is conducted while varying the sample sizes over the range given by $N \in \{20, 200, 1000\}$. In this simulation, the anchor points, which must be provided a priori for basis function generation, are simply given by 3 by 3 points (see Figure 8). In the second simulation, the influence of the number of anchor points on the predictive accuracy is examined using a fixed sample size of N=200 synthetic samples. There, the numbers of regularly placed anchor points assumed are 3×3, 5×5, 7×7, 10×10, and 13×13. In the first and second simulations, predictive accuracy evaluation is iterated over 200 replicated experiments. These simulations are performed using 64 bit PC whose memory is 4.0 GB.

---

[1] Based on Eq.(25), $Bw_i$ can be considered as a rough estimate of $f_i$, which is determined by the monitored temperatures and not depend on hot-tweets. Use of the rough estimate is important to increase the computational efficiency.

### 4.2.2 Results of accuracy and efficiency Monte Carlo studies

Table 5 summarizes root mean squared errors (RMSEs) of the two approaches when the sample size is increasing from N= 20, 200, and 1000 samples respectively. This table clearly shows the inaccurateness of the basis function approach relative to the kernel-based widely applied approach for a fixed number of basis functions given by 3×3. Even though the number of samples increases substantially the basis function approach is still underperforming in this case. We believe it is primarily due to the fact that bias due to small numbers of anchor points in the representation is resulting in this reduced accuracy. We will test this hypothesis in the second study by fixing the number of samples and increasing the number of anchor nodes.

Table 6 summarizes the RMSE with variation of the number of anchor points for a moderate number of sample observations N=200. This table suggests that, while basis function approach is less accurate when the number of anchor points is small, the accuracy is compatible with the kernel-based approach when the number of anchor points is large. This result suggests the importance of using a sufficient number of anchor points, and, accordingly, basis functions.

Clearly the basis function approach is significantly more computationally efficient than the kernel based method, due to its dimension reduction. Hence, when the appropriate number of basis anchor points are selected, this can be highly computational efficient. To see this, computational times of the two approaches are summarized in Table 7. This figure demonstrates that our basis function approach is far more computationally efficient than commonly used kernel-based approach, in particular, when the sample size is large.

In short, the basis function approach is a computationally efficient approach whose accuracy is compatible provided that a sufficient number of basis functions are used. Based on the result, the next sub-section applied the basis function approach to temperature estimation while paying attention to how to determine the number of basis functions used.

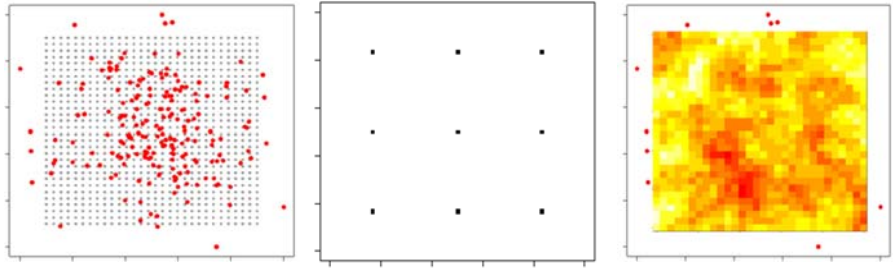

**Fig. 8.** Examples of sample sites and 30 by 30 regular pointes (left), 3 by 3 anchor points (middle), and a Gaussian process generated (right).

**Table 5.** Sample size and RMSE

| Sample size | Kernel-based approach | Basis function approach (3 by 3 anchor points) |
|---|---|---|
| 20 | 1.047 | 1.820 |
| 200 | 0.707 | 1.778 |
| 1000 | 0.399 | 1.251 |

**Table 6.** The number of anchor points and RMSE (sample size: 200)

| # of anchor points | Kernel-based approach | Basis function approach |
|---|---|---|
| 3×3 |  | 1.778 |
| 5×5 |  | 1.157 |
| 7×7 | 0.707 | 0.853 |
| 10×10 |  | 0.813 |
| 13×13 |  | 0.784 |

**Table 7.** Computational times (second)

| Sample size | Kernel-based approach | Basis function approach (3 by 3 anchor points) |
|---|---|---|
| 20 | 0.30 | 0.00 |
| 200 | 20.7 | 0.12 |
| 1000 | 1124.51 | 1.08 |

## 4.3. Application to temperature interpolation

### 4.3.1 Assumptions

This section applies S-BLUE Eq.(19) for a spatiotemporal temperature interpolation in Tokyo in August 25, 2012. Because computationally efficiency is crucially important in real-time risk management, which is our focus, we apply the basis function approach for the covariance function estimation.

The calculation procedure is as follows: (i) hourly mean temperature, which captures time trends of spatially distributed temperatures, and distance to the Tokyo station and distance to the nearest station, which capture the heat island effect, are regressed on the monitored temperatures; (ii) the spatiotemporal process of the residual temperatures is modeled by the basis function approach; (iii) temperatures are interpolated by S-BLUE with hot-tweets (S-BLUEwith) and S-BLUE without hot-tweets (S-BLUEwithout). S-BLUEwith utilizes both hourly monitored temperature data at 8 monitoring stations (Figure 1) and hot-tweets (see Figure 3), and S-BLUEwithout utilizes only the monitored temperature data. Their interpolation equations are   and  , respectively. Section 4.3.2 selects the number of basis functions, which is critical for accurate interpolation (see the previous section), and section 4.3.3 discusses the interpolation result.

### 4.3.2 Selection of the number of anchor points and basis function

This section selects the number of anchor points by a 5-fold cross-validation (CV) that quantifies the model accuracy by the following steps: (i) divide the monitored temperature data into 5 subsamples randomly; (ii) apply 4/5 subsamples for the model estimation; (iii) predicts the remaining 1/5 subsamples using the estimated model; (iv) evaluate the predictive accuracy by comparing monitored and predicted temperatures using RMSE; (v) repeat steps (ii) to (iv) for all 5 cases. The 5-fold CV is conducted iteratively by varying the number of anchor points.

Since we consider both spatial and temporal dimensions, anchor points must be placed in both of these dimensions. Regarding the 1D temporal dimension, At anchor points are placed at regular intervals, and At is varied during the CV. Regarding the 2D spatial dimension, we place anchor points at each of the 8 monitoring stations, and they are fixed throughout the analysis.

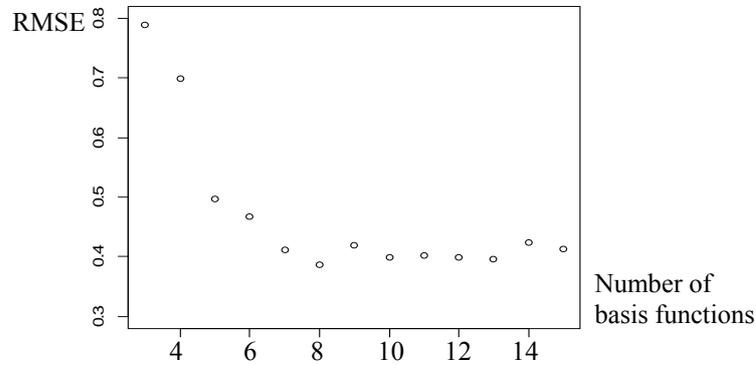

**Fig. 9.** RMSE and number of basis functions

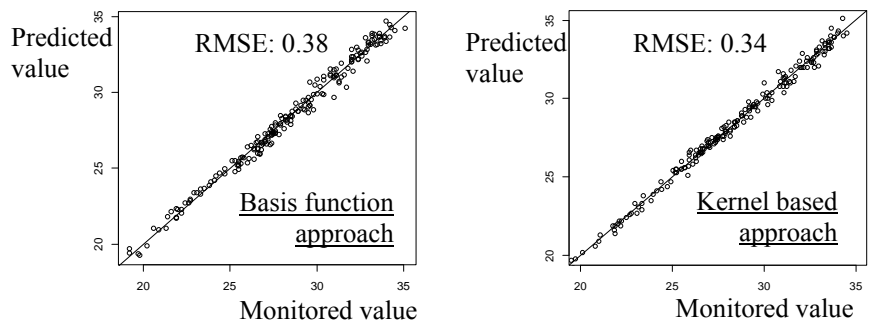

**Fig. 10.** Comparison of monitored and predicted temperatures: 8 basis functions are applied in the basis function approach.

Figure 9 displays the relationship between RMSE and the number of basis functions. This figure suggests that, while it is crucial to introduce more than a contain number of basis functions, RMSE is nearly constant once the number of basis function goes beyond 8. Hence, we apply 8 basis functions, which reduces the computational cost and provides the most stable, accurate and parsimonious model choice in this study. Note that the 8 basis function case furnishes the smallest RMSE. Figure 10, which compares monitored and predicted temperatures, shows that accuracy of the basis function-based approach is comparable to the kernel-based approaches. This figure confirms the computational efficiency of the basis function approach for large spatiotemporal data (e.g., twitter data) descriptions.

### *4.3.3 Interpolation result*

Figure 11 plots the interpolated temperatures at 6AM, 0PM, and 6PM in August 25, 2012. A significant difference between S-BLUE$_{with}$ and S-BLUE$_{without}$ are found at 0PM: while the northern area is the hottest based on S-BLUE$_{without}$, the center of Tokyo area, where hot-tweets are densely distributed at around 0PM, is the hottest based on S-BLUE$_{with}$. Considering the severe heat island effect in the central Tokyo area (e.g., Saitoh et al., 1996; Ohashi et al., 2007), the latter result is intuitively more reasonable.

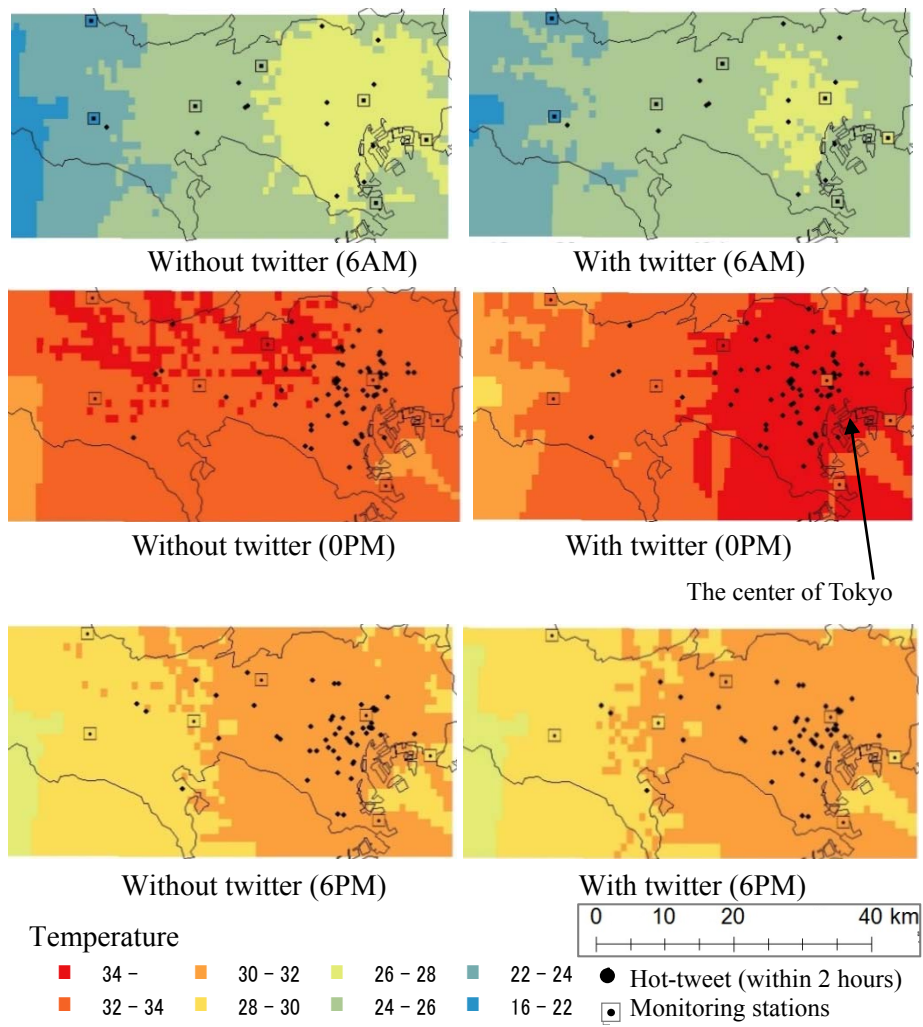

**Fig. 11.** Temperature interpolation results

Accuracy of S-BLUE$_{with}$ and S-BLUE$_{with}$ out are compared by a cross-validation. Since consideration of hot-tweets must be significant for temperature interpolation at unmonitored sites, temperatures monitored at 7 stations (and hot-tweets) are applied to interpolate temperatures at the remaining station site, and it is iterated for all 8 cases. Then, predictive accuracy is evaluated by RMSE.

RMSE of S-BLUE$_{with}$ and S-BLUE$_{without}$ are 3.85 and 3.89 only at the monitoring sites respectively, verifying that consideration of tweets increase local temperature interpolation accuracy marginally. Though due to the calibration of the models at these sites, it is not unreasonable to expect similar performance. However, more importantly, as shown in Figure 11, we see that the incorporation of the geo-tagged Twitter data can significantly change the estimated spatial resolution accuracy for local area analysis. This finding would be valuable as a first step of utilizing twitter and other social medias in urban climate analysis. We believe this is the first step in the incorporation of participatory sensing data with ground based sensor, as with all first attempts, this can be improved also with remote satellite sensing data. We believe the study of the twitter data and the approach of extracting hotness-related tweets can be enhanced further in future analysis.

Figure 12 displays RMSEs evaluated in every monitoring sites. The figure suggests that the consideration of hot-tweets significantly increases the accuracy at two monitoring stations in the bayside part of the central area. The improvement suggests that densely distributed hot-tweets in the central area (see Figure 12) successfully captures temperature-related information that could not be captured by monitored temperatures. Based on Figure 12, the improvement might be due to the fact that hot-tweets were required to capture heat island effects in the bayside area around noon accurately.

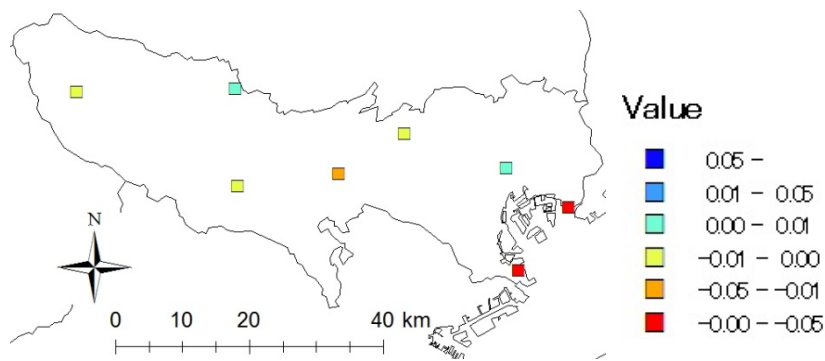

**Fig. 12.** Gap of RMSEs (S-BLUE$_{with}$ minus S-BLUE$_{without}$)

## 5. Concluding remarks

This study first revealed that hot-tweets increase depending on not only temperatures but also change of temperatures. In addition, we showed that rivers and parks decrease perceived hotness, and hot-tweets as well. After that, as an application example, we applied the twitter data for a local temperature interpolation, and showed that tweets are beneficial to increase the accuracy.

Our discussion suggests the potential of Twitter in real-time management of temperature-related risks, including the risk of urban heatwave. Considering the trend of global warming and the fact that some past heatwaves have caused many deaths, real-time risk management of heatwave must be an urgent task. Fortunately, tweets reflect not only ambient temperatures, but also conditions of each human. A risks analysis using both temperature information and human condition information in tweets (participatory sensing data) would be an important study toward a real-time urban heatwave management utilizing social media data.

# References

An, X., Ganguly, A.R., Fang, Y. (2014) Tracking climate change opinions from twitter data. Workshop on Data Science for Social Good held in conjunction with KDD 2014.

Bollen, J., Mao, H., Zeng, X-J. (2011) Twitter mood predicts the stock market. Journal of Computational Science, 2 (1), 1–8.

Cressie, N. (1993) Statistics for Spatial Data. Wiley.

Cressie, N., Johannesson, G. (2008) Fixed rank kriging for very large spatial data sets. Journal of the Royal Statistical Society. Series B (Statistical Methodology), 70 (1), 209–226.

Cressie, N., Wikle, C.K. (2013) Statistics for Spatiotemporal Data. Wiley.

Cruz, M.G., Peters, G.W., Shevchenko P.V. (2014) Fundamental Aspects of Operational Risk and Insurance Analytics: A Handbook of Operational Risk. John Wiley & Sons.

Diener, A., O'Brein, B., Gafni, A. (1998) Health care contingent valuation studies: a review and classification of the literature. Health Economics, 7 (4), 313–326.

Hahmann, S., Purves, R.S., Burghardt, D. (2014) Twitter location (sometimes) matters: Exploring the relationship between georeferenced tweet content and nearby feature classes. Journal of Spatial Information Science, 9, 1–36.

Hoff, P.D., Niu. X. (2012) A Covariance Regression Model. Statistica Sinica, 22, 729–753.

Kauermann, G., Schellhase, C., Ruppert, D. (2013) Flexible Copula Density Estimation with Penalized Hierarchical B-Splines. Scandinavian Journal of Statistics, 40 (4), 685–705.

Li, R., Lei, K.H., Khadiwala, R., Chang, K.C-C. (2012) TEDAS: a twitter based event detection and analysis system. Data Engineering (ICDE), 123–1276.

Nakamichi, K., Yamagata, Y., Seya, H. (2013) $CO_2$ emissions evaluation considering introduction of EVs and PVs under land-use scenarios for climate change mitigation and adaptation–Focusing on the change of emission factor after the Tohoku Earthquake–. Journal of the Eastern Asia Society for Transportation Studies, 10, 1025–1044.

Nevat I, Peters, GW, Collings, I.B. (2013) Random field reconstruction with quantization in wireless sensor networks. Signal Processing, IEEE Transactions on, 61 (23), 6020–6033.

Nevat, I, Peters, G.W., Septier, F., Matsui, T. (2015) Estimation of spatially correlated random fields in heterogeneous wireless sensor networks. Signal Processing, IEEE Transactions on, 63 (10), 2597–2609.

Ohashi, Y., Genchi, Y., Kondo, H., Kikegawa, Y., Yoshikado, H., Hiraon, Y. (2007) Influence of air-conditioning waste heat on air temperature in Tokyo during summer: Numerical experiments using an urban canopy model coupled with a building energy model. Journal of Applied Meteorology and Climatology, 46, 66–81.

Peters, G.W., Matsui, T. (2015) Modern Methodology and Applications in Spatial-Temporal Modeling, Springer.

Rosen, S. (1974). Hedonic prices and implicit markets: Product differentiation in pure competition. Journal of Political Economy, 82 (1), 34–55.

Saitoh, T.S., Shimada, T., Hoshi, H. (1996) Modeling and simulation of the Tokyo urban heat island. Atmospheric Environment, 30 (20), 3431–3442.

Sklar, A. (1959) Fonctions de répartition à n dimensions et leurs marges. Publications de l'Institut de Statistique de l'Université de Paris, 8, 229–231.

Thelwall, M., Buckley, K., Paltoglou, G. (2011) Sentiment in twitter events. Journal of The American Society for Information Science and Technology, 62 (2), 406–418.

Wood, S.N. (2003) Thin plate regression splines. Journal of the Royal Statistical Society B, 65 (1), 95–114.

Wood, S.N. (2006) Generalized Additive Models: An Introduction with R. Chapman & Hall/CRC Press, New York.

Wood, S.N., Scheipl, F., and Faraway, J.J. (2013) Straightforward intermediate rank tensor product smoothing in mixed models. Statistics and Computing, 23 (3), 341–360.

Yamagata, Y., Seya, H. (2013) Simulating a future smart city: An integrated land use-energy model. Applied Energy, 112, 1466–1474.

Zhang, J. (2010) Multi-source remote sensing data fusion: status and trends. International Journal of Image and Data Fusion, 1 (1), 5–24.